# Determination of the force transmission error in a single-sinker magnetic suspension densimeter due to the fluid-specific effect and its correction for use with gas mixtures containing oxygen

Daniel Lozano-Martín[1], María E. Mondéjar[2], José J. Segovia[1], and César R. Chamorro[1,*].

[1] Grupo de Termodinámica y Calibración (TERMOCAL), Research Institute on Bioeconomy, Escuela de Ingenierías Industriales. Universidad de Valladolid. Paseo del Cauce, 59, 47011 Valladolid, Spain.

[2] DTU Mechanical Engineering, Technical University of Denmark, Nils Koppels Allé, Building 403, 2800 Kgs. Lyngby, Denmark.

**Abstract**

Density measurements from single-sinker magnetic suspension densimeters need to be corrected to compensate for the magnetic effects of the measuring cell materials and the fluid on the coupling transmission system. While the magnetic effect of the densimeter materials can be easily determined, the fluid effect requires the calculation of an apparatus-specific constant, $\varepsilon_\rho$. In this work, the apparatus-specific constant of the single-sinker magnetic suspension densimeter at the University of Valladolid has been determined by using two alternative methods. The first method, which uses density data for the same fluid and conditions and different sinkers, yielded a value of $\varepsilon_\rho = 4.6 \cdot 10^{-5}$. The second method, obtained from measurements with pure oxygen, yielded a value of $\varepsilon_\rho = 8.822 \cdot 10^{-5}$. The second value is considered as more reliable, as the first method presents inherent limitations in this case.

* Corresponding author e-mail: cescha@eii.uva.es. Tel.: +34 983423756. Fax: +34 983423363

1. **Introduction**

Single-sinker magnetic suspension densimeters (SSMSD) provide very precise density measurements for fluids over extensive temperature and pressure limits [1]. In their operation, a sinker of known volume, immersed in the fluid whose density must be determined, is weighed by an analytical balance. The density of the fluid is determined from the buoyancy force on the sinker, according to the Archimedes' principle. The sinker, surrounded by the fluid at the desired temperature and pressure, is coupled to the balance, which unlike the sinker is in air at ambient conditions, across the wall of the measuring cell by means of a contactless magnetic suspension coupling. A comprehensive description of the measuring method can be found in [2].

The balance reading when weighing the sinker through the magnetic suspension coupling is influenced by the magnetic properties of the materials of the measuring cell and by the magnetic properties of the fluid. This is accounted as a force transmission error (FTE) that must be evaluated and corrected for [3] for single-sinker densimeters and gravimetric sorption analyzers and other scientific equipment using a magnetic-suspension coupling [4].

There are different ways of estimating the correction that must be applied to compensate for the FTE. Kuramoto et al. [5] proposed a theoretical and quantitative determination of the FTE by developing a physical model, whereas Kano et al. [6] studied the FTE by using numerical methods, with the finite element analysis. These two methods require a detailed knowledge of the magnetic properties of all the materials in the cell, the sinker and the fluid, together with a precise knowledge of the geometry of the magnetic coupling and all the surrounding elements. This information is not always easily accessible, especially when the relative position of the magnetic coupling changes with temperature and pressure, due to the change of the magnetic properties of the permanent magnet with the temperature. Based on these works, Kayukawa et al. [7] developed a dual-sinker magnetic suspension densimeter capable of keeping the vertical position of the permanent magnet constant during the entire measuring process, which minimizes the influence of the FTE. However, the mechanical complexity of this approximation is only of interest for some very accurate metrological issues and not practical for most of the SSMSD.

Alternatively, McLinden et al. [3] proposed an empirical approach. The FTE can be split into two terms: the term due to the magnetic behavior of the apparatus materials (e.g. cell walls, sinker) and the fluid-specific term. The apparatus-specific term of the FTE can be easily determined by measuring the weight of the sinker in the evacuated cell. The fluid-specific term of the FTE can be neglected when dealing with diamagnetic fluids, which have weak magnetic susceptibility, but must be considered when measuring densities of paramagnetic fluids, which have a strong and temperature-dependent magnetic susceptibility, such as mixtures containing oxygen. The fluid-specific term of the FTE, as well as the apparatus-specific effect, are unique for each SSMSD, and therefore, must be determined independently for each densimeter.

The SSMSD at the University of Valladolid (UVa) [8][9] has been used during the last few years to measure the density of different binary mixtures [8,10–16] and multicomponent mixtures [17–20]. As no significant amount of any paramagnetic component was present in those mixtures, only the apparatus-specific correction was considered. The fluid-specific term of the FTE was neglected and its small influence was considered to be covered by the uncertainty of the density measurements. A new set of density measurements is scheduled to be performed with this SSMSD for binary mixtures of relevance for processes involved in carbon capture and storage systems. Some of the relevant mixtures contain significant amounts of oxygen [21], and therefore the determination of the fluid-specific term of the FTE is needed in order to correct these measurements.

In this work we have followed the procedure proposed by McLinden et al. [3] for the determination of the fluid-specific term of the FTE of the SSMSD at the UVa. The fluid-specific term of the FTE, as it will be explained in the next section, is proportional to the magnetic susceptibility and the density of the fluid, with the proportionality being constant the so-called 'apparatus-specific constant', $\varepsilon_\rho$. In order to avoid confusions arising from the use of the language, it is worth highlighting that the apparatus-specific constant, $\varepsilon_\rho$, is the proportionality constant of the fluid-specific term of the FTE, and should not be mistaken with the apparatus-specific term of the FTE.

The apparatus-specific constant, $\varepsilon_\rho$, can be evaluated by measuring the density of a fluid at the same temperature and pressure using two different sinkers. For this purpose, any fluid with known magnetic susceptibility is appropriate, even if the fluid is diamagnetic (i.e. with small and negatives values of the magnetic susceptibility). An alternative approach to estimate the apparatus-specific constant is by measuring the density of a paramagnetic fluid, such as pure oxygen, at selected temperatures and pressures. In this case the magnetic susceptibility is much greater and with a positive value. In this work we have followed the two different approaches to estimate $\varepsilon_\rho$ and the results are compared.

## 2. Force transmission error in a single-sinker magnetic suspension densimeter

The operation of the SSMSD is schematically depicted in Figure 1. The magnetic coupling is formed by an electromagnet hanging from the lower hook of the analytical balance and a permanent magnet inside the measuring cell. The permanent magnet is fixed to a sinker support which allows coupling and decoupling the sinker to/from the balance. The magnetic coupling has two different positions: the zero position (ZP) and the measuring position (MP). A position sensor allows changing from one position to the other.

A load compensation system permits running the analytical balance near its zero, avoiding any systematic errors associated with the non-linearity of the balance. Additionally, the load compensation system allows calibrating the balance at each measuring point. The load compensation system consists of two calibrated masses. Usually, one of them is made out of tantalum and the other of titanium. They have distinct masses but nearly the same volume, thus the air buoyancy effect on the compensation masses are the same for both of them. The difference in weight between the two masses is designed to be similar to the sinker mass.

In the ZP the magnetic coupling attracts the permanent magnet but the sinker is not lifted and it rests on the bottom of the cell. In the MP the magnetic coupling attracts the permanent magnet so that the sinker is lifted. Simultaneously, the load compensation system places the tantalum mass on the upper pan of the balance when the magnetic coupling is in the ZP, and places the titanium mass when the magnetic coupling is in the MP. The balance readings in the ZP, $W_{ZP}$, and in the MP, $W_{MP}$, are therefore:

$$W_{ZP} = \alpha\{\Phi[m_{pm} - \rho_{fluid}V_{pm}] + m_{em} + m_{Ta} - \rho_{air}(V_{em} + V_{Ta})\} \quad (1)$$

$$W_{MP} = \alpha\{\Phi[m_s + m_{pm} - \rho_{fluid}(V_s + V_{pm})] + m_{em} + m_{Ti} - \rho_{air}(V_{em} + V_{Ti})\} \quad (2)$$

where the factor $\alpha$ accounts for the balance calibration; $\Phi$ is the coupling factor, which corrects for the effect of the magnetic coupling in the reading of the balance; $m$, $V$ and $\rho$, denote mass, volume and density, respectively; the subscripts *pm*, *em*, *s*, refer to the permanent magnet, the electromagnet and the sinker; and the subscripts *Ta* and *Ti*, refer to the tantalum and titanium compensation masses.

The difference between these two readings is

$$W_{MP} - W_{ZP} = \alpha\{\Phi[m_s - \rho_{fluid}V_s] + m_{Ti} - m_{Ta} - \rho_{air}(V_{Ti} - V_{Ta})\} \quad (3)$$

The differential nature of the measuring method cancels the weights of the permanent magnet, the electromagnet, and their corresponding buoyancy forces. As the volumes of the titanium and tantalum compensation loads are approximately equal ($V_{Ti} \approx V_{Ta}$), the air buoyancy term for them is negligible. The density of the fluid, $\rho_{\text{fluid}}$, can be isolated from equation (3), thus obtaining the following expression:

$$\rho_{\text{fluid}} = \rho_{\text{s}} + \frac{1}{\Phi} \frac{(m_{\text{Ti}}-m_{\text{Ta}})\mp(W_{\text{ZP}}-W_{\text{MP}})/\alpha}{V_{\text{s}}} \qquad (4)$$

where $\rho_{\text{s}}$ is the density of the sinker ($\rho_{\text{s}} = m_{\text{s}}/V_{\text{s}}$). Equation (4) gives the density of the fluid, but includes the unknown balance calibration factor, $\alpha$, and the coupling factor, $\Phi$.

The balance calibration factor, $\alpha$, can be easily obtained by an independent calibration, free of the air buoyancy effect, using the two compensation masses. As the magnetic coupling is in the zero position (ZP), two measurements are taken, one with the titanium mass on the balance pan ($W_{\text{ZP,Ti}}$), and the other one with the tantalum mass ($W_{\text{ZP,Ta}}$). Subtracting one from the other most of the terms in equation (1) cancel and $\alpha$ can be obtained:

$$\alpha = \frac{W_{\text{ZP,Ta}}-W_{\text{ZP,Ti}}}{(m_{\text{Ta}}-m_{\text{Ti}})-\rho_{\text{air}}(V_{\text{Ta}}-V_{\text{Ti}})} \qquad (5)$$

The analysis of the coupling factor for a two-sinker MSD performed by McLinden et al. [3] states that, the coupling factor, $\Phi$, can be separated into an apparatus-specific effect, $\Phi_0$ and a fluid-specific effect $\Phi_{\text{fse}}$, as expressed in equation (6). This is also valid for a SSMSD.

$$\Phi = \Phi_0 + \Phi_{\text{fse}} \qquad (6)$$

The apparatus-specific effect, $\Phi_0$, can be obtained by weighing the sinker in a vacuum. In that case $\rho_{\text{fluid}} = 0$ and $\Phi_{\text{fse}} = 0$, and thus, $\Phi_0$ can be isolated from equation (4) as:

$$\Phi_0 = \frac{-(m_{\text{Ti}}-m_{\text{Ta}})\pm(W_{\text{ZP,vacuum}}-W_{\text{MP,vacuum}})/\alpha}{m_{\text{s}}} \qquad (7)$$

It is worth noting that the apparatus-specific effect, $\Phi_0$, depends on temperature, because the strength of the permanent magnet is temperature-dependent, and therefore the equilibrium position of the magnetic coupling changes for each temperature. Furthermore, small changes on the alignment of the electromagnet, or in the leveling of the balance, also produce small variations in the value of $\Phi_0$. For this reason, the value of $\Phi_0$ must be estimated at the end of each single isotherm and the corresponding correction must be done with the obtained value.

From the analysis for the two-sinker MSD by McLinden et al. [3], it was also demonstrated that the fluid-specific effect is proportional to the magnetic susceptibility and density of the fluid:

$$\Phi_{\text{fse}} = \varepsilon_\rho \frac{\chi_{\text{s}}}{\chi_{\text{s0}}} \frac{\rho_{\text{fluid}}}{\rho_0} \qquad (8)$$

where $\chi_{\text{s0}} = 10^{-8}$ m$^3\cdot$kg$^{-1}$ and $\rho_0 = 1000$ kg$\cdot$m$^{-3}$ are reducing constants for the specific magnetic susceptibility of the fluid, $\chi_{\text{s}}$, and the density of the fluid, $\rho_{\text{fluid}}$, respectively; and $\varepsilon_\rho$ is a so-called 'apparatus-specific constant'.

If the correction due to the FTE is not considered, i.e., setting $\Phi = 1$ in Equation (4), the density obtained, $\rho_{\Phi=1}$, has a deviation from the 'true' value. The relative deviation in density is given by the following expression (equation 23 from McLinden et al. [3]):

$$\frac{\Delta\rho}{\rho_{\text{fluid}}} = \frac{\rho_{\Phi=1}-\rho_{\text{fluid}}}{\rho_{\text{fluid}}} = (\Phi - 1)\left(1 - \frac{\rho_{\text{s}}}{\rho_{\text{fluid}}}\right) \qquad (9)$$

Replacing the value of $\Phi$ given by equations (6) and (8) in equation (9), the following is obtained:

$$\frac{\Delta\rho}{\rho_{\text{fluid}}} = (\Phi_0 - 1)\left(1 - \frac{\rho_s}{\rho_{\text{fluid}}}\right) + \varepsilon_\rho \frac{\chi_s}{\chi_{s0}}\left(\frac{\rho_{\text{fluid}}}{\rho_0} - \frac{\rho_s}{\rho_0}\right) \quad (10)$$

The first term on the right-hand side of equation (10) stands for the relative deviation in density due to the apparatus effect, which will be easily compensated for with the appropriate value of $\Phi_0$ obtained from equation (7). The second term represents the relative deviation in density due to the fluid-specific effect:

$$\frac{\Delta\rho_{\text{fse}}}{\rho_{\text{fluid}}} = \varepsilon_\rho \frac{\chi_s}{\chi_{s0}}\left(\frac{\rho_{\text{fluid}}}{\rho_0} - \frac{\rho_s}{\rho_0}\right) \quad (11)$$

As it can be seen in equation (11), the relative deviation in density due to the fluid-specific term of the FTE is proportional to the specific magnetic susceptibility of the fluid, $\chi_s$, and to the difference between the density of the fluid and the density of the sinker, $\rho_{\text{fluid}} - \rho_s$, being the proportionality constant that is the apparatus specific constant, $\varepsilon_\rho$. McLinden et al. [3] showed that the apparatus-specific constant, $\varepsilon_\rho$, can be calculated from equation (11) following two different approaches. The first method is by measuring the density of a fluid at exactly equal $T$ and $p$, using two different sinkers, which for a single-sinker MSD implies the disassembling and re-assembling of the whole cell. The second method is by measuring the density of pure oxygen, a strongly paramagnetic fluid with a high magnetic susceptibility. Both approaches have been used in this work to estimate the apparatus-specific constant, $\varepsilon_\rho$, of the SSMSD at the UVa.

The volume magnetic susceptibility is a dimensionless magnitude. The magnetic susceptibility, $\chi_s$, that appears in equation (11) refers to the magnetic susceptibility on a *mass* basis (in $m^3 \cdot kg^{-1}$). The product of the specific magnetic susceptibility by the density, which appears in equation (11), gives the dimensionless volume magnetic susceptibility. Most of the values of magnetic susceptibilities in the literature are given as molar magnetic susceptibilities ($\chi_M$, in $m^3 \cdot mol^{-1}$). The specific magnetic susceptibility, $\chi_s$, can be obtained easily by dividing the molar magnetic susceptibilities, $\chi_M$, by the molar mass, $M$, of the substance. In some cases, the published data of magnetic susceptibility are given in the centimeter-gram-second system of units (CGS-EMU). In this case, a convenient conversion factor, which can be found in Ref. [22], should be used to obtain the magnitude in the corresponding SI units.

Once the value of the apparatus-specific constant, $\varepsilon_\rho$, is established, the density of the fluid can be easily obtained by introducing the value of the coupling factor, $\Phi$, in equation (4), obtaining the following expression:

$$\rho_{\text{fluid}} = \frac{\Phi_0 m_s + (m_{\text{Ti}} - m_{\text{Ta}}) \mp (W_{\text{ZP}} - W_{\text{MP}})/\alpha}{V_s} \frac{1}{\Phi_0} + \frac{\varepsilon_\rho}{\Phi_0} \frac{\chi_s}{\chi_{s0}}\left(\frac{\rho_s}{\rho_0} - \frac{\rho_{\text{fluid}}}{\rho_0}\right)\rho_{\text{fluid}} \quad (12)$$

where an approximate value of $\rho_{\text{fluid}}$ must be introduced in the right-hand side of the equation to obtain the experimental value of $\rho_{\text{fluid}}$. The value of density given by equation (4) with $\Phi = \Phi_0$ after more than one interaction, is a good approximation for this purpose. Otherwise, the value of $\rho_{\text{fluid}}$ computed from RefProp [23] can be used.

### 3. Experimental

### 3.1. The single-sinker magnetic suspension densimeter at UVa

The SSMSD at UVa was described thoroughly in the papers by Chamorro et al. [8] and by Mondéjar et al. [9]. Formerly, the SSMSD had a cylindrical titanium sinker, with a nominal mass of 60 g and a nominal volume of 13 cm$^3$. The 'old' titanium sinker was replaced by a 'new' monocrystalline silicon cylinder, with nearly the same mass but double the volume (26 cm$^3$), as the density of silicon is approximately half that of titanium. This modification was made in order to reduce the measuring uncertainty in density measurements and increase the resolution of the SSMSD, mainly at low densities. Table 1 collects the mass, volume and density of the 'old' titanium sinker and the 'new' silicon sinker of the SSMSD at UVa. The measuring cell is made of a copper-chromium-zirconium diamagnetic alloy (CuCrZr) with $\chi_s/\chi_{s0}$ = - 0.025 [24].

The temperature and pressure dependence of the volume of the sinker must be considered when using equation (4) or (12) to estimate the density of the fluid. This can be done by using equation (13):

$$V_s(T,p) = V_{s0}\left(1 + 3\overline{\alpha(T)}(T - T_0) - \frac{1}{K(T)}(p - p_0)\right) \tag{13}$$

where $V_{s0}$ stands for the volume of the sinker at the reference state, specified in its calibration certificate; $\overline{\alpha(T)}$ is the average linear thermal expansion coefficient; $K(T)$ is the compressibility modulus, which is more frequently expressed as function of the Young's modulus, $E$, and Poisson's ratio, $v$, as $K(T) = 3[1 - 2v(T)]/E(T)$; and $T_0$ and $p_0$ are the reference state temperature and pressure. The expression for the dependence of the thermal expansion coefficient with temperature has been obtained from [25] for the titanium sinker and from [26,27] for the silicon sinker. The corresponding expressions for dependence of the mechanical properties with the temperature have been obtained from [28]. Note that, as the silicon is an anisotropic material, the average of the elastic constants for each crystalline direction have been considered.

A high-precision analytical balance (Mettler Toledo XPE205DR, Mettler Toledo GmbH, Gießen, Germany, normal weighing range: 81 g, readability: 0.01 mg). measures the buoyancy force on the sinker through the magnetic coupling system. The load compensation system is formed by two calibrated masses, one of them made out of tantalum and the other titanium. The calibrated compensation masses have approximately the same volume (4.9 cm$^3$) and the difference in weight between both of them is similar to the sinker mass (approx. 60 g for the 'old' titanium sinker and 62 g for the 'new' silicon sinker). The two masses were provided by Rubotherm GmbH, Bochum, Germany. Their masses and volumes were calibrated at the Spanish National Metrology Institute (Centro Español de Metrología, CEM). The mass, volume and density of the two calibrated masses of the load compensation system (tantalum and titanium) are presented in Table 1.

The temperature is set by an outer double-walled stainless-steel cylinder, through which the fluid from a Julabo FP50 thermal bath circulates, and an inner electrical heating cylinder made of copper, directly in contact with the measuring cell and powered by an electronic Julabo MC-E controller. Two platinum resistance thermometers (S1059PJ5X6, Minco Products, Inc., Minneapolis MN, USA) located one opposite the other in the middle section of the measuring cell and plugged into an AC comparator resistance bridge (F700, Automatic Systems Laboratories, Redhill, England) measure the temperature of the gas inside the cell. The probes have been calibrated on the ITS-90 scale [29]. Two pressure transducers, a Paroscientific 2500A-101 (0 to 3 MPa) and a Paroscientific 43KR-HHT-101 (Paroscientific Inc., Redmond WA, USA), for pressures up to 20 MPa, measure the gas pressure. These pressure transducers are frequently calibrated against a dead weight pneumatic pressure balance and load masses with traceability to international standards.

## 4. Results

### 4.1. Determination of the apparatus-specific constant using the single-sinker method with two sinkers of different density

The apparatus-specific constant, $\varepsilon_\rho$, can be estimated by measuring the density of a fluid of known magnetic susceptibility, $\chi_s$, at the same $T$ and $p$, using, at different times, the same SSMSD with two different sinkers of distinct density [3]. Following this method, density measurements of nitrogen performed with the original setup of the SSMSD at the UVa (with the 'old' titanium sinker) and with the current configuration of the SSMSD (with the 'new' monocrystalline silicon sinker) can be used. It must be noted that the original objective of those measurements was not to determine the apparatus-specific constant, but simply to check the performance of the SSMSD with a reference fluid.

For this purpose, equation (11) can be used twice, with the results of the measurements of nitrogen performed with the SSMSD with the two different sinkers, to evaluate the relative deviation in the density of pure nitrogen due to the fluid-specific effect in both cases:

$$\frac{\Delta\rho_{\text{fse},1}}{\rho_{\text{fluid},1}} = \varepsilon_\rho \frac{\chi_s}{\chi_{s0}} \left( \frac{\rho_{\text{fluid},1}}{\rho_0} - \frac{\rho_{s,1}}{\rho_0} \right) \tag{14}$$

$$\frac{\Delta\rho_{\text{fse},2}}{\rho_{\text{fluid},2}} = \varepsilon_\rho \frac{\chi_s}{\chi_{s0}} \left( \frac{\rho_{\text{fluid},2}}{\rho_0} - \frac{\rho_{s,2}}{\rho_0} \right) \tag{15}$$

where $\rho_{\text{fluid},1}$ and $\rho_{\text{fluid},2}$ are the experimental density given by the densimeter with the sinker 1 and the sinker 2, respectively. $\rho_{s,1}$ and $\rho_{s,2}$ are the densities of the 'old' titanium sinker (sinker 1) and the 'new' silicon sinker (sinker 2), indicated in Table 1.

If measurements are done at nearly the same $T$ and $p$ ($\rho_{\text{fluid},1} \approx \rho_{\text{fluid},2}$), the value of $\varepsilon_\rho$ can be obtained from equations (14) and (15), by subtracting one from the other:

$$\varepsilon_\rho = \frac{\chi_{s0}}{\chi_s} \left( \frac{\rho_0}{\rho_{s,2} - \rho_{s,1}} \right) \left( \frac{\Delta\rho_{\text{fse},1} - \Delta\rho_{\text{fse},2}}{\rho_{\text{fluid}}} \right) \tag{16a}$$

If the measurements with the sinker 1 and the sinker 2 are made at different conditions, the values of pressure and temperature may be slightly different for each measurement (more so in this case, as the measurements were not made for the purpose of being compared in mind). This can be solved by comparing the experimental results with an appropriate equation of state at the experimental temperature and pressure and adapting equation (16a) to this modification, as shown below:

$$\frac{\Delta\rho_{\text{fse},1} - \Delta\rho_{\text{fse},2}}{\rho_{\text{fluid}}} = \frac{\rho_{1,\Phi=\Phi_0} - \rho_{\text{EoS}}(T_1,p_1)}{\rho_{\text{EoS}}(T_1,p_1)} - \frac{\rho_{2,\Phi=\Phi_0} - \rho_{\text{EoS}}(T_2,p_2)}{\rho_{\text{EoS}}(T_2,p_2)} \tag{16b}$$

where the subscripts 1 and 2 stand for the measurements with the sinker 1 or 2, respectively, and $\rho_{\text{EoS}}$ is the estimated density from the equation of state for nitrogen of Span et al [30].

Nitrogen is a diamagnetic fluid with a molar magnetic susceptibility of $\chi_M = -0.151 \cdot 10^{-9}$ m$^3 \cdot$mol$^{-1}$ at $T = 293.15$ K [24,31]. Taking $M_{\text{N2}} = 28.01348$ g$\cdot$mol$^{-1}$ as the molar mass of nitrogen, this gives a specific magnetic susceptibility of nitrogen of $\chi_s = 5.39 \cdot 10^{-9}$ m$^3 \cdot$kg$^{-1}$. The temperature and pressure dependence of $\chi_s$ for diamagnetic fluids is negligible [22].

Therefore, selected nitrogen density data measured with the old titanium sinker were compared with the density data of pure nitrogen measured with the new silicon sinker. The quality factors of the used nitrogen are summarized in Table 2. Table 3 gives the ($p$, $\rho$, $T$) values of the two sets of density measuring points used for the estimation of $\varepsilon_\rho$. The values of $\varepsilon_\rho$ calculated with equations (16a and 16b) for each pair of measuring points are also presented in Table 3 and plotted versus density in Figure 2.

As it can be observed, the scatter of the $\varepsilon_\rho$ values is significantly high at low densities and is reduced at high densities. Values of the apparatus specific constant yielded an average value of $\varepsilon_\rho = 4.6 \cdot 10^{-5}$, but showed a significant scattering. It must be noted that the value of the apparatus-specific constant following this single-sinker method with two sinkers of different density has been evaluated with very few density data, and not specifically obtained for this purpose. The obtained value is not very reliable, as a very small variation in the density data, even within the small experimental uncertainty of the equipment, yields important changes in the value of $\varepsilon_\rho$.

Using the obtained value of $\varepsilon_\rho$, the value of density for pure nitrogen, corrected for the fluid-specific effect, can be obtained through equation (12). The correction due to the fluid-specific effect will be very small as the magnetic susceptibility of nitrogen is very small, and results in a correction of less than 0.01 % in density, for a nominal value of $\rho = 300$ kg·m$^{-3}$.

### 4.2. Determination of the apparatus-specific constant by density measurements on pure oxygen

McLinden et al. [3] stated that the apparatus specific constant, $\varepsilon_\rho$, may also be evaluated by measuring the density of pure oxygen at near ambient conditions ($T = 293.15$ K and $p = 0.1$ MPa). This method, simpler than the single-sinker method with two sinkers of different density, will give an acceptable approximation of the apparatus-specific constant. If the densimeter allows the measurement of pure oxygen at higher pressures and different temperatures (considering material compatibility with oxygen and important safety related issues) it will provide more accurate results.

Equation (11) could then be used to obtain $\varepsilon_\rho$:

$$\varepsilon_\rho = \phi_0 \left[ \frac{\rho_{\Phi=\Phi_0} - \rho_{EoS}}{\rho_{EoS}} \right] / \left[ \frac{\chi_s}{\chi_{s0}} \left( \frac{\rho_{EoS}}{\rho_0} - \frac{\rho_s}{\rho_0} \right) \right] \tag{17}$$

where $\rho_{\Phi=\Phi_0}$ is the density given by the densimeter, corrected for the apparatus-specific effect but not for the fluid-specific effect, and $\rho_{EoS}$ is the density given by a reference equation of state at the experimental temperature and pressure.

The multiparameter equation of state for non- and weakly polar fluids, particularized for oxygen, proposed by Span and Wagner [32] has been used as the reference equation of state for oxygen. The molar magnetic susceptibility of molecular oxygen at the reference state, $T = 293.15$ K and $p = 0$ MPa, is taken as $\chi_{M00} = (42.92 \pm 0.06) \cdot 10^{-9}$ m$^3$·mol$^{-1}$, value measured by May et al. [33], and in close accordance with the ab initio calculation of Minaev et al. [34]. Taking $M_{O2} = 31.9988$ g·mol$^{-1}$ as the molar mass of oxygen, this gives a specific magnetic susceptibility of oxygen at the reference state of $\chi_{00} = 1.34130 \cdot 10^{-6}$ m$^3$·kg$^{-1}$. This value is 250 times larger than the absolute value of the specific magnetic susceptibility of nitrogen. The dependence of the magnetic susceptibility of strong paramagnetic fluids, as oxygen, with temperature should not be neglected [33]. The Curie Law, which states that the magnetic susceptibility is proportional to $1/T$, has been used to account for this temperature dependence:

$$\chi_s = \chi_{00} \frac{293.15\ K}{T} \tag{18a}$$

Even when the magnetic susceptibility of pure oxygen may also have a weak dependence with density, decreasing the magnetic susceptibility with increasing density, this effect is quite small [33] and has not been considered initially.

Therefore, density measurements on pure oxygen were performed with the SSMSD at the UVa at (250, 273.15, 293.15, 325, 350, and 375) K and pressures up to 6 MPa. The measurement of oxygen at high pressures required a thorough review of material compatibility for our SSMSD [35] and other safety related issues [36]. The quality parameters of the pure oxygen are summarized in Table 2. Measurements were performed along each isotherm starting from the higher pressure, and decreasing pressure by 1 MPa steps to the lower pressure. Thirty points were measured for each temperature and pressure to make sure equilibrium had been reached, and the mean value of the last ten measurements was used to estimate $\varepsilon_\rho$.

The fourth column of Table 4 and the corresponding data shown in Figure 3 show the relative deviations of the obtained values of density of oxygen corrected for the apparatus-specific effect but not for the fluid-specific effect, $\rho_{\Phi=\Phi_0}$, from the density given by the reference equation of state for oxygen by Span and Wagner [32], $\rho_{EoS}$. It can be seen that the deviations range from 2.1 % for the density data at 375 K to 3.1 % for the density data at 250 K. There is an appreciable dependence on temperature, with the deviations being larger at lower temperatures and decreasing with increasing temperatures. The dependence of the deviations with pressure is less significant.

The ninth column of Table 4 and Figure 4 show the values obtained for $\varepsilon_\rho$, by using equation (17). It can be seen that the values of $\varepsilon_\rho$ ranges from $8.4 \cdot 10^{-5}$ (for $T = 250$ K, $p = 5.7$ MPa, and $\rho = 91.8$ kg·m$^{-3}$) to $9.2 \cdot 10^{-5}$ (for $T = 350$ K, $p = 1.0$ MPa, and $\rho = 10.7$ kg·m$^{-3}$). The obtained values of $\varepsilon_\rho$ increase with temperature and decrease with density. The excellent repeatability of $\varepsilon_\rho$ for each temperature and pressure can also be seen in Figure 4, where the last ten values of $\varepsilon_\rho$ for each temperature and pressure have been represented in the graph and not only its mean value, which is reflected in Table 4. It can be seen that the last ten measurements for each single $(T, p)$ point show a very little scatter, although it increases slightly as the density decreases.

The values of $\varepsilon_\rho$ were adjusted to a linear equation as a function of $T$ and $\rho$ separately. The resulting equation for $\varepsilon_\rho$, as a function of temperature and density, is:

$$\varepsilon_\rho(T,\rho) = 8.822 \cdot 10^{-5} + 4.698 \cdot 10^{-8} \cdot (T/K - 293.15) - 3.015 \cdot 10^{-8} \cdot \rho/(kg \cdot m^{-3})$$
(19a)

The residuals between the experimental values of $\varepsilon_\rho$ and the values given by equation (19a) are presented graphically in Figure 5. The adjustment of equation (19a) to the experimental data is very good, with a root-mean-square deviation, expressed as a percentage, of less than 0.44 % ($4.01 \cdot 10^{-7}$).

Using the value of $\varepsilon_\rho$ given by equation (19a), the value of density for pure oxygen, corrected for the fluid-specific effect, can be obtained through equation (12). The sixth column of Table 4 and Figure 6 show the relative deviations of the obtained values of density of oxygen corrected for the apparatus-specific effect and for the fluid-specific effect, $\rho_{fluid}$, from the density given by the reference equation of state for oxygen by Span and Wagner [32], $\rho_{EoS}$. It can be seen that all the experimental density data are located within the claimed uncertainty of the equation of state (0.2 %). The average absolute deviation (*AAD*) of the experimental data with respect to this equation of state is 0.009 %, with a maximum deviation of 0.012 %, as shown in Table 5.

Comparison of the experimental density results were also made with the reference equation of state for oxygen by Schmidt and Wagner [37], which has a lower uncertainty. The results are presented in the seventh column of Table 4 and in Figure 7. It can be seen that all the experimental density data lay within the stated uncertainty of the equation of state (0.1 %). The average absolute deviation (*AAD*) of the experimental data with respect this equation of state is 0.023 %, with a maximum deviation of 0.026 %.

The results of the statistical analysis of the comparison of the experimental density data against these two reference equations of state for oxygen are presented in Table 5.

In the case that the slight dependence of the magnetic susceptibility of pure oxygen with the density is also considered, as indicated in [33], and not only the dependence with temperature, equation (18a) should be rewritten as:

$$\chi_s = \frac{1}{M}\left(\frac{N_A\mu_0\left(\alpha_m + \mu_m^2/3k_BT\right)(1+b_\mu\rho)}{1-\rho\frac{N_A\mu_0}{3}\left(\alpha_m + \mu_m^2/3k_BT\right)(1+b_\mu\rho)}\right) \tag{18b}$$

where $N_A$ stands for the Avogadro's constant, $\mu_0 = 4\pi \cdot 10^{-7}$ N·A$^{-2}$, $\alpha_m$ stands for the molecular magnetizability, $\mu_m$ stands for the molecule's magnetic dipole moment, $k_B$ stands for the Boltzmann constant, and $b_\mu$ stands for the second magnetic virial coefficient. For oxygen, the recommended values are [33]: $N_A\mu_0\alpha_m = -0.13 \cdot 10^{-9}$ m$^3$·mol$^{-1}$, $\mu_m = 2.6282 \cdot 10^{-23}$ N·m·T$^{-1}$, and $b_\mu = -1.8 \pm 0.5$ cm$^3$·mol$^{-1}$.

With this magnitude for the magnetic susceptibility, the value obtained for $\varepsilon_\rho$ from the experimental densities of oxygen, by using equation (17), will be slightly different to equation (19a):

$$\varepsilon_\rho(T,\rho) = 8.822 \cdot 10^{-5} + 4.801 \cdot 10^{-8} \cdot (T/K - 293.15) - 2.522 \cdot 10^{-8} \cdot \rho/(kg \cdot m^{-3}) \tag{19b}$$

As can be seen in equation (19b), even when the dependence of the magnetic susceptibility with density is considered, there is still an appreciable dependence of the value of $\varepsilon_\rho$ with density. Comparing equation (19b) to equation (19a) the coefficient for $\rho$ decreases only by 16 %, from $3.015 \cdot 10^{-8}$ to $2.522 \cdot 10^{-8}$, being the coefficient for $T$ and the independent term of the equation practically the same in both equations.

Using the value of $\varepsilon_\rho$ given by equation (19b), the value of density for pure oxygen, corrected for the fluid-specific effect, can also be obtained through equation (12), but in this case, the value for the magnetic susceptibility, corrected for temperature and density (equation 18b), should be used. Obviously, the results are exactly the same as when equation 12 is used with the values of $\varepsilon_\rho$ given by equation (19a) and the magnetic susceptibilities given by equation (18a).

### 4.3. Estimation of the uncertainty of the apparatus-specific constant and its influence on the uncertainty of density.

The uncertainty of each single value of $\varepsilon_\rho$ obtained for a selected temperature and pressure from density measurements on pure oxygen can be obtained by applying the law of propagation of uncertainty [38] to equation (17). For this purpose, the contributions from the uncertainty of the experimental density, corrected for the apparatus-specific effect but not for the fluid-specific effect, $U(\rho_{\Phi=\Phi0})$, the uncertainty of the reference EoS from Span and Wagner, $U(\rho_{EoS})$, the uncertainty of the specific magnetic susceptibility of O$_2$, $U(\chi_s)$, and the uncertainty of the density of the silicon sinker, $U(\rho_s)$, must be considered.

The uncertainty of the experimental density, corrected for the apparatus-specific effect but not for the fluid-specific effect, $U(\rho_{\Phi=\Phi0})$, is given by the expression $U(\rho_{\Phi=\Phi0})/kg \cdot m^{-3} = 1.1 \cdot 10^{-4} \cdot \rho/kg \cdot m^{-3} + 2.3 \cdot 10^{-2}$, as it was explained in detail in [9]. The uncertainty in density of the reference EoS from Span and Wagner [32], $U(\rho_{EoS})$, is less than 0.2 %. The uncertainty of the specific magnetic susceptibility of $O_2$, $U(\chi_s)$, is obtained from [33] and accounts to up to 0.14 %. Finally, the uncertainty of the density of the silicon sinker, $U(\rho_s)$, is of 0.014 %, as indicated by its calibration certificate from the Spanish National Metrology Institute, CEM.

From these values, the estimated relative expanded uncertainty ($k = 2$) of $\varepsilon_\rho$ amounts to 4 % for higher densities, but increases up to 20 % for lower densities. Under the assumption of normal probability density functions, the relative uncertainty of the extrapolated values to zero density of $\varepsilon_\rho$ have been computed by the Monte Carlo method [39] from a linear regression as function of $\rho$, with an average result of 0.97 %.

The uncertainty of the experimental densities of oxygen, corrected by both the apparatus-specific and the fluid-specific FTE effects, $U(\rho_{fluid})$, is straightforward evaluated from the application of the law of propagation of uncertainty to equation (12). The relative expanded ($k = 2$) uncertainty of the density increases from 0.08 % at high densities (for $\rho > 95$ kg·m$^{-3}$), up to 0.43 % for low densities (for $\rho \approx 10$ kg·m$^{-3}$). The estimated relative uncertainty for the 293.15 K isotherm is depicted graphically with error bars in Figures 6 and 7. The uncertainty of the experimental density, $U(\rho)$, considering the effect of the uncertainty of $\varepsilon_\rho$, can be given, as an approximation, by the expression:

$$U(\rho)/\text{kg·m}^{-3} = 2.5 \cdot 10^4 \cdot \chi_s/\text{m}^3 \cdot \text{kg}^{-1} + 1.1 \cdot 10^{-4} \cdot \rho/\text{kg·m}^{-3} + 2.3 \cdot 10^{-2} \qquad (20)$$

## 5. Discussion

The scatter of the $\varepsilon_\rho$ values obtained from nitrogen density measurements by using the single-sinker method with two sinkers of different density is significantly high, mainly at low densities. Small variations in the density data, even within the experimental uncertainty of the equipment, results in significant changes in the value of $\varepsilon_\rho$, giving even negative values for $\varepsilon_\rho$, as can be seen in Figure 2, which has no physical meaning. The mean value obtained for $\varepsilon_\rho$ by using the single-sinker method with two sinkers of different density is $4.6 \cdot 10^{-5}$, but this value is very unreliable, as it has been calculated with very few nitrogen density data, not specifically obtained for this purpose. In order to obtain a better result for $\varepsilon_\rho$ following this single-sinker method with two sinkers of different density, more density data should be used, preferably with fluids with a higher magnetic susceptibility, higher density and with two sinkers with a bigger difference in density (both with respect each other and with respect the density of the fluid measured), as can be deducted from the main influence parameters in equation (16a). For example, this may be achieved using a diamagnetic gas of higher susceptibility, as methane ($\chi_s/\chi_{s0}$ = -1.36), or even better synthetic air ($\chi_s/\chi_{s0}$ = 30.07), and two sinkers, one of silicon and the other of tantalum instead of titanium ($\rho_{Ta} \approx 7.2\rho_{Si}$), which would lead to a contribution of the fluid-specific effect about 0.21 % according to equation (11) (or about -5.1 % in case of synthetic air), rather than the 0.01 % contribution when using the silicon and titanium sinkers ($\rho_{Ti} \approx 1.9\rho_{Si}$) and nitrogen ($\chi_s/\chi_{s0}$ = -0.54). In this way, the effect would be below achievable relative uncertainties of density measurements of methane not worse than 0.03 %, and the single-sinker method with two sinkers of different density may be appropriate, thus, yielding more accurate results of $\varepsilon_\rho$. The apparatus-specific constant is unique for each SSMSD and should be recalculated when a major modification of the equipment is performed [3,4,40]. This is an inherent limitation of the single-sinker method with two sinkers of different density: the apparatus-specific constant should be slightly different for the same SSMSD with two different sinkers, and this method provides only one single value. The value obtained by the single-sinker method with two sinkers of different density is, therefore, a mean value of two slightly different $\varepsilon_\rho$, one for the SSMSD with each different sinker.

The value for $\varepsilon_\rho$ obtained by measuring the density of pure oxygen has a weak, but appreciable, dependence on temperature and density. The value of $\varepsilon_\rho$ ranges from $8.4 \cdot 10^{-5}$ (for $T = 250$ K, $p = 5.7$ MPa, and $\rho = 91.8$ kg·m$^{-3}$) to $9.2 \cdot 10^{-5}$ (for $T = 350$ K, $p = 1.0$ MPa, and $\rho = 10.7$ kg·m$^{-3}$). The value of $\varepsilon_\rho$ as a function of $T$ and $\rho$, can be obtained from equation (19). This dependence of $\varepsilon_\rho$ with $T$ and $\rho$ can be explained because the geometry of the magnetic coupling changes with temperature and density. The influence of the temperature in the fluid-specific effect of the FTE is equivalent to the influence detected in the apparatus-specific effect, $\Phi_0$, recorded in the literature [3,4,40]. This effect of temperature in the value of $\varepsilon_\rho$, due to the change in the geometry of the magnetic coupling, is avoided by some authors [6] by changing the measurement procedure, making the magnetic coupling always work at the same distance between electromagnet and permanent magnet. But also, the magnetic susceptibility of the measuring cell changes with the temperature. Although the housing or our equipment is made of a diamagnetic alloy of CuCrZr, the presence of other paramagnetic parts or even undesirable impurities of ferromagnetic particles in the surroundings of the magnetic coupling may shift the overall magnetic behavior from diamagnetic to paramagnetic. This is our case, as it is reflected from the ($\Phi_0$-1) values obtained when measuring in vacuum, with a strong temperature dependence, with results changing from negative to positive values, ranging from $-13.2 \cdot 10^{-6}$ at $T = 250$ K, up to $2.8 \cdot 10^{-6}$ at $T = 375$ K.

The value of $\varepsilon_\rho$ obtained in this work by this second method is double the value obtained by the first method. However, the mean value obtained for $\varepsilon_\rho$ with the first method, if the negative values of $\varepsilon_\rho$ (which have no physical meaning) and the outliers (two of the experimental $\varepsilon_\rho$ are beyond the 75 % quartile) are eliminated, is of $8.4 \cdot 10^{-5}$, and this value is close to the value given by the second method. Anyway, this value is of no significance, as it is obtained from only 12 experimental densities of nitrogen measured with two different sinkers. Additionally, with so few pairs of experimental density data, the first method is unable to detect any dependence of $\varepsilon_\rho$ with temperature. Even when McLinden [3] states that the second method will give only an approximation of the fluid-specific effect correction, in our case this second method has proven to be more precise. This is mainly because we have measured the density of pure oxygen not only at a single point, near ambient conditions ($T = 293.15$ K and $p = 0.1$ MPa), but over a wide range of temperatures and pressures.

For all these reasons, we will take the value of $\varepsilon_\rho$ obtained by the second method as the reference value for the corrections due to the fluid-specific effect on the FTE for the SSMSD at UVa. This value is of the same order of magnitude as the values of $\varepsilon_\rho$ given by several authors for similar SSMSD's. $\varepsilon_\rho = 18.9 \cdot 10^{-5}$ for the densimeter at Texas A&M University in Texas, USA [40], $\varepsilon_\rho = 3.6 \cdot 10^{-5}$ and $\varepsilon_\rho = 4.4 \cdot 10^{-5}$ (before and after an important modification of the magnetic suspension coupling) for the SSMSD at the Ruhr- Universität in Bochum, Germany [3].

## 6. Conclusions

Two different approaches were used to estimate the apparatus-specific constant, $\varepsilon_\rho$, of the fluid-specific term of the force transmission error of the magnetic-suspension coupling of the single-sinker magnetic suspension densimeter at the University of Valladolid. The numerical results obtained are specific for the SSMSD at UVa, but, from a qualitative point of view, the conclusions obtained in this work are valid for other SSMSD and other scientific equipment using a magnetic-suspension coupling.

The mean value obtained for $\varepsilon_\rho$ by using a single-sinker method with two sinkers of different density is $4.6 \cdot 10^{-5}$, but this value has a large uncertainty, as it was calculated with very few nitrogen density data, not specifically obtained for this purpose. The value for $\varepsilon_\rho$ obtained by measuring the density of pure oxygen has a weak, but appreciable, dependence on temperature and density. The value of $\varepsilon_\rho$ as a function of $T$ and $\rho$, can be obtained from equation (19). The value of $\varepsilon_\rho$ obtained by this second method is double the value obtained with the first method. Nevertheless, both values are compatible, given the large uncertainty of the value obtained by the first method. For these reasons, the value of $\varepsilon_\rho$ obtained by the second method should be considered as more reliable.

The correction due to the fluid-specific effect will be very small when measuring densities of diamagnetic fluids (very low magnetic susceptibility) and when using sinkers of density close to the density of the fluid (low density sinkers), as can be deducted from equation (11). In the case of the SSMSD at UVa, with the new silicon sinker (with a relatively low density), the correction due to the fluid-specific effect result in a correction of less than 0.01 % in density, for a nominal value of $\rho = 300$ kg·m$^{-3}$, for a nitrogen sample, or for any other gas sample of diamagnetic fluids. Therefore, the omission of the fluid specific effect correction for mixtures of diamagnetic fluids is acceptable if maximum accuracy is not required.

When measuring mixtures with high oxygen content, which will imply a high magnetic susceptibility of the mixture, namely up to two orders of magnitude larger than diamagnetic fluids, the correction due to the fluid-specific effect should be considered, even with SSMSD with low-density sinkers installed.

**Acknowledgements**

The authors want to thank Ministerio de Economía, Industria y Competitividad project ENE2017-88474-R and Junta de Castilla y León project VA280P18 for their support.

**Figures**

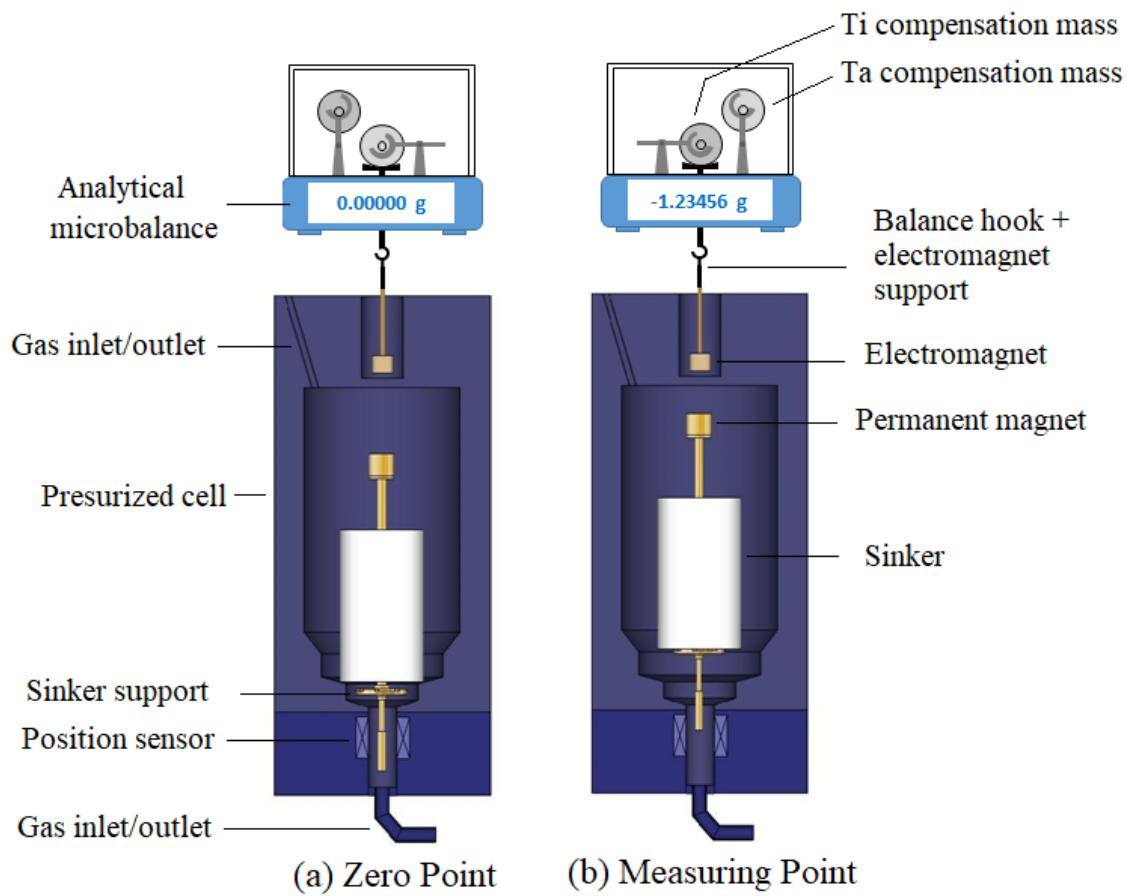

**Figure 1.** Schematic diagram of the single-sinker magnetic suspension densimeter, SSMSD: (a) Zero Point, and (b) Measuring Point.

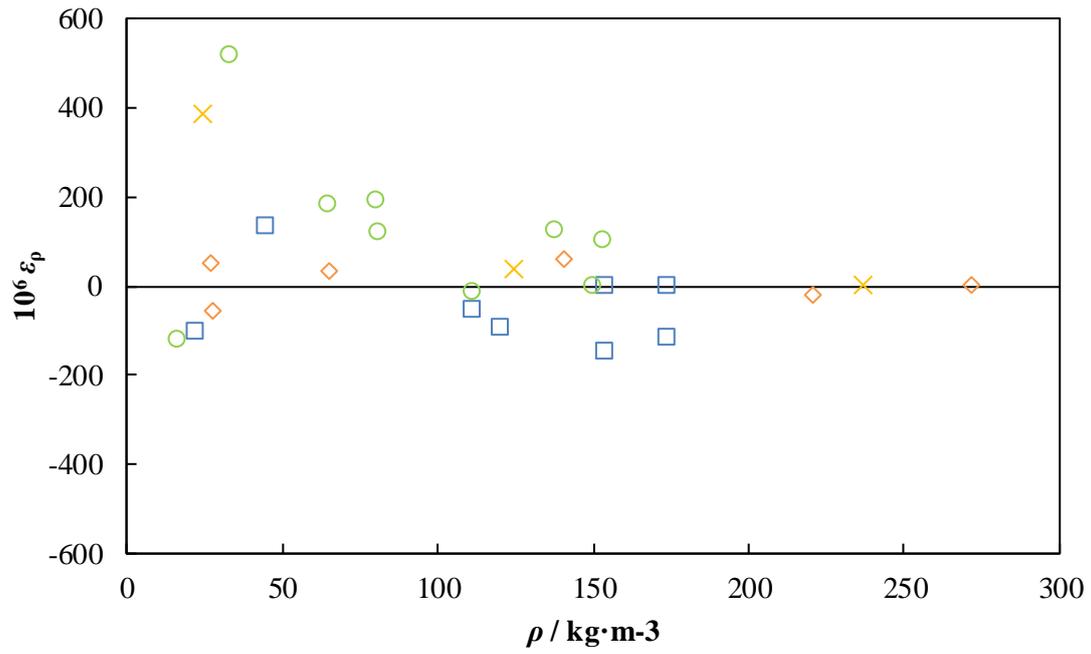

**Figure 2.** Apparatus-specific constant, $\varepsilon_\rho$, calculated from pairs of nitrogen densities measured at nearly the same temperature and pressure conditions with the old titanium sinker (sinker 1) and the new silicon sinker (sinker 2), at four different nominal temperatures: ◇ 250 K, × 275 K, □ 300 K, ○ 400 K.

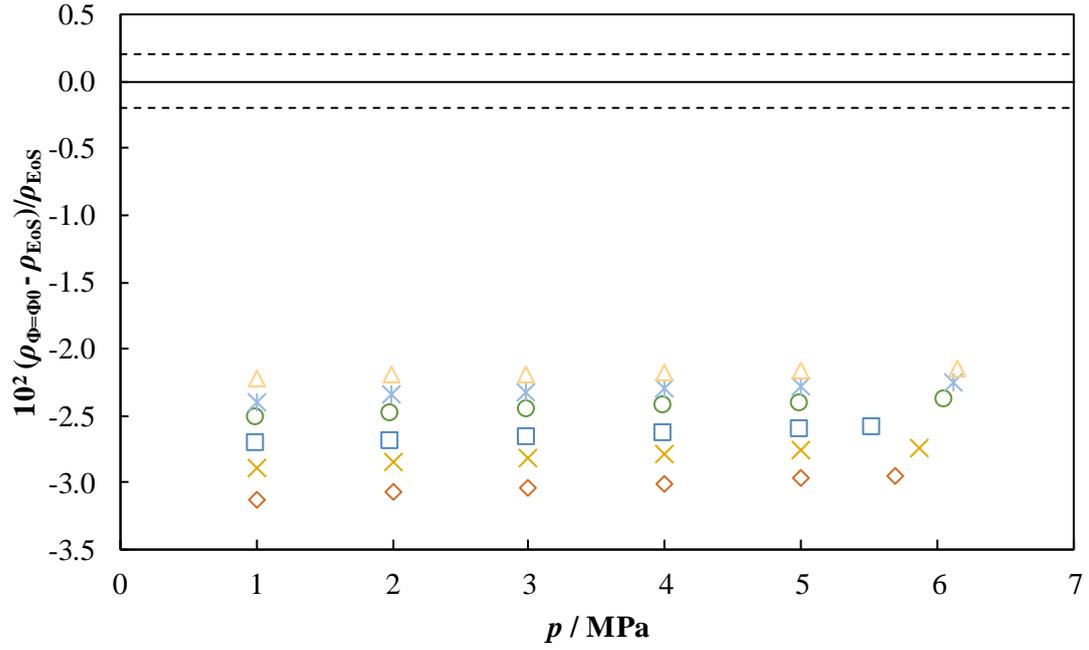

**Figure 3.** Relative deviations, $100 \cdot (\rho_{\Phi=\Phi_0} - \rho_{EoS})/\rho_{EoS}$, of the experimental values of density of oxygen corrected solely for the apparatus-specific effect (but not for the fluid-specific effect), $\rho_{\Phi=\Phi_0}$, from the density given by the reference equation of state for oxygen by Span and Wagner [32], $\rho_{EoS}$, as a function of pressure for different temperatures: ◇ 250 K, × 273.15 K, □ 293.15 K, ○ 325 K, ✻ 350 K, △ 375 K. Dashed lines indicate the expanded ($k = 2$) uncertainty of the Span and Wagner EoS of 0.2 %.

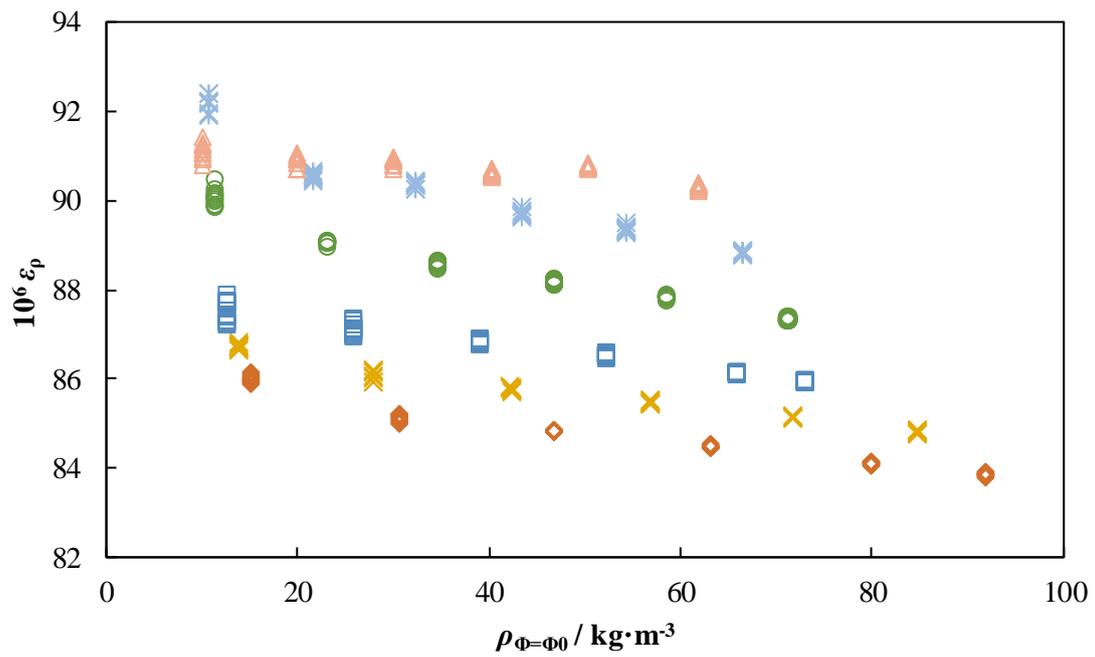

**Figure 4.** Values obtained for the apparatus-specific constant, $\varepsilon_\rho$, by using equation (17) at different temperatures and pressures as a function of the experimental density: ◇ 250 K, × 273.15 K, □ 293.15 K, ○ 325 K, ✶ 350 K, △ 375 K.

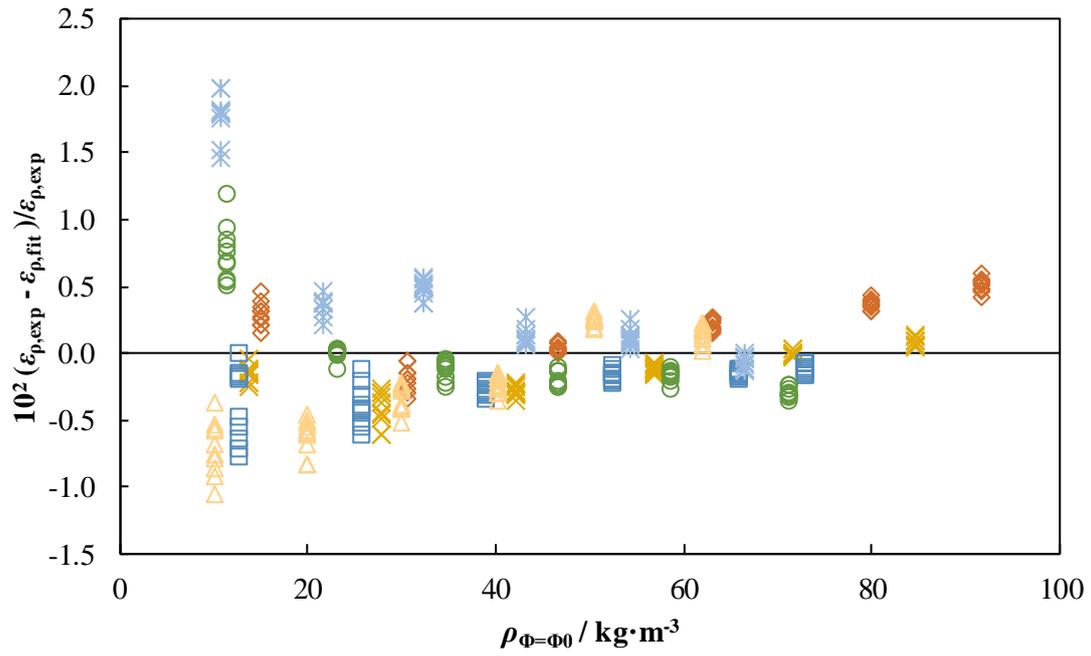

**Figure 5.** Residual analysis ($\varepsilon_{\rho,exp} - \varepsilon_{\rho,fit})/\varepsilon_{\rho,exp}$ as a function of density of the measured $\varepsilon_{\rho,exp}$ and the values fitted $\varepsilon_{\rho,fit}$ by equation (19a): ◇ 250 K, × 273.15 K, □ 293.15 K, ○ 325 K, ✶ 350 K, △ 375 K.

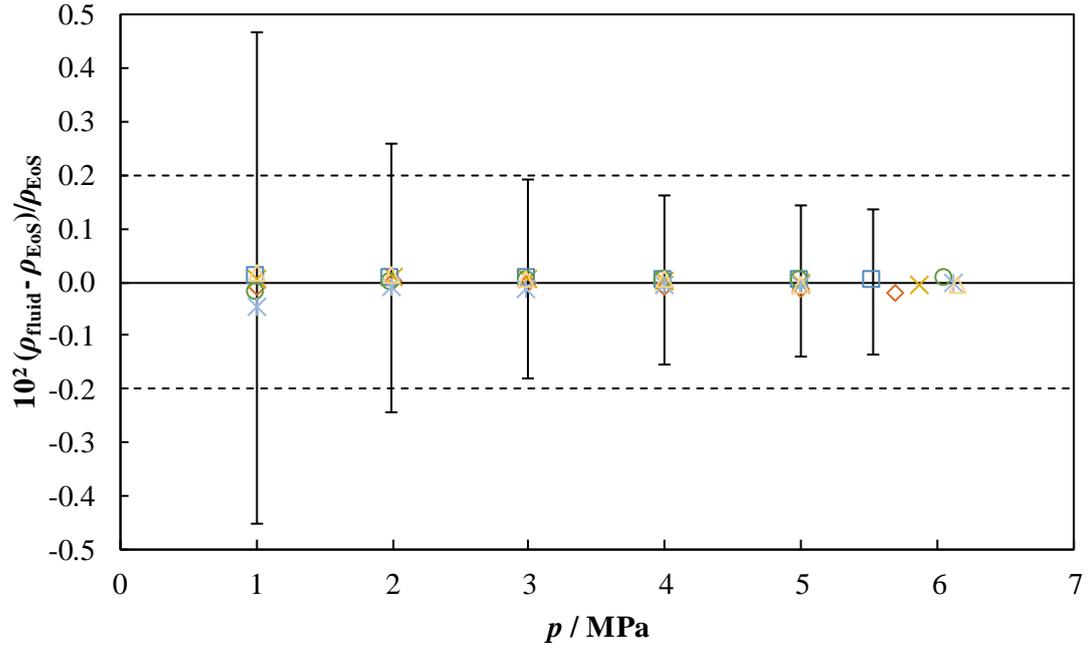

**Figure 6.** Relative deviations, $(\rho_{fluid} - \rho_{EoS})/\rho_{EoS}$, of the experimental values of density of oxygen corrected for both the apparatus-specific effect and the fluid-specific effect, $\rho_{fluid}$, from the density given by the reference equation of state for oxygen by Span and Wagner [32], $\rho_{EoS}$, as a function of pressure for different temperatures: ◇ 250 K, × 273.15 K, □ 293.15 K, ○ 325 K, ✷ 350 K, △ 375 K. Dashed lines indicate the expanded ($k = 2$) uncertainty of the Span and Wagner EoS of 0.2 %. Error bars (only given for the experimental densities at 293.15 K) indicate the expanded ($k = 2$) uncertainty of the experimental density.

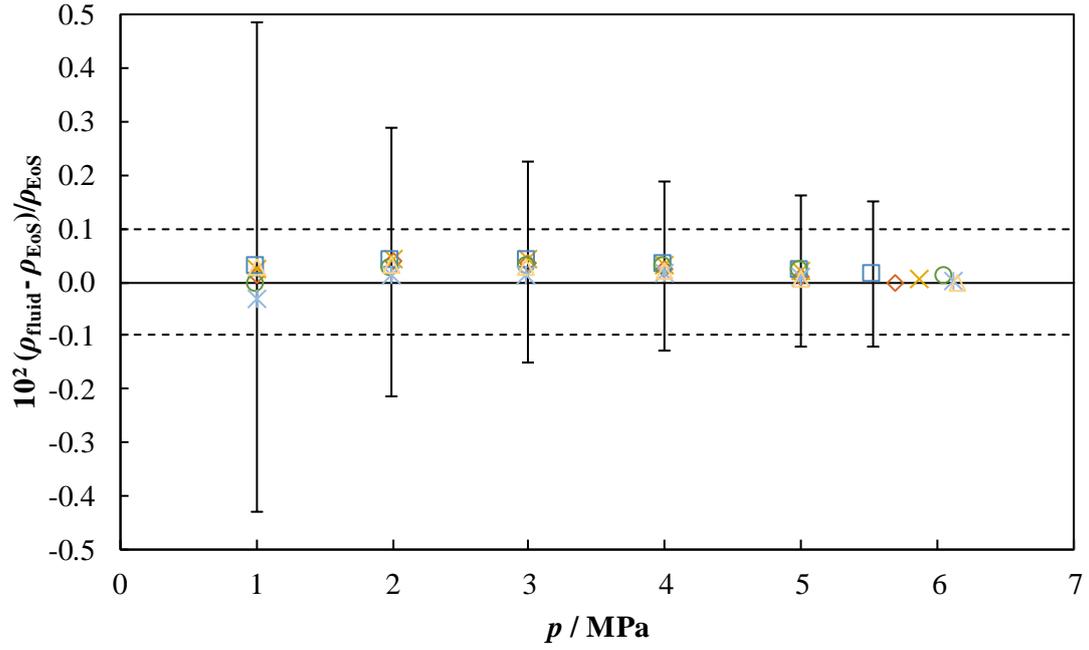

**Figure 7.** Comparison of the experimental density results of oxygen corrected for both the apparatus-specific effect and the fluid-specific effect, $\rho_{\text{fluid}}$, with the reference equation of state for oxygen by Schmidt and Wagner [37], $\rho_{\text{EoS}}$, as a function of pressure for different temperatures: ◇ 250 K, × 273.15 K, □ 293.15 K, ○ 325 K, ✶ 350 K, △ 375 K. Dashed lines indicate the expanded ($k = 2$) uncertainty of the Schmidt and Wagner EoS of 0.1 %. Error bars for the 293.15 K isotherm indicate the expanded ($k = 2$) uncertainty of the experimental density.

**Tables**

**Table 1.** Specifications of the two load compensation masses and the two different sinkers of the SSMSD at the University of Valladolid (UVa).

| | Mass[a] / g | Volume[b,d] / cm$^3$ | Density[c,d] / kg·m$^{-3}$ |
|---|---|---|---|
| Tantalum mass (load compensation system) | 82.0883 | 4.9240 | 16670.9 |
| Titanium mass (load compensation system) | 22.39968 | 4.9706 | 4506.5 |
| 'Old' titanium sinker (sinker 1) | 59.68962 | 13.2648 | 4506.5 |
| 'New' silicon sinker (sinker 2) | 61.59181 | 26.4440 | 2329.12 |

[a] $10^6\ U(m)/m = 2.0$

[b] $10^6\ U(V_0)/V_0 = 100$

[c] $10^6\ U(\rho_0)/\rho_0 = 140$

[d] At ambient conditions of $t_{air}$ = 20.8 °C, $t_{water}$ = 19.965 °C, $p$ = 947.19 mbar, and Relative Humidity = 53.1 %

**Table 2.** Purity, supplier, molar mass, and critical parameters for the pure gases used in this work.

|  | Purity / mol-% | Supplier | $M$ / g·mol$^{-1}$ | Critical parameters[a] | |
|---|---|---|---|---|---|
|  |  |  |  | $T_c$ / K | $P_c$ / MPa |
| Nitrogen | 99.9999 | Air Liquide | 28.0135 | 126.19 | 3.3958 |
| Oxygen | 99.9999 | Linde | 31.9988 | 154.58 | 5.043 |

[a] Critical parameters were retrieved for each substance from RefProp software [23]

**Table 3.** Nitrogen ($p$, $\rho$, $T$) values of the measuring points compared to the density given by the reference equation of state for nitrogen by Span et al [30]. The subscripts 1 and 2 stands for the measures with sinker 1 ('old' titanium sinker) or 2 ('new' silicon sinker), respectively. Values of $\varepsilon_\rho$ calculated with equation (16) for each pair of measuring points.

| $T_1$ / K[a] | $p_1$ / MPa[a] | $\rho_{1,\Phi=\Phi_0}$ / kg·m$^{-3}$ | $10^6$ ($\rho_{1,\Phi=\Phi_0}$ - $\rho_{Span}$)/$\rho_{Span}$ | $T_2$ / K[a] | $p_2$ / MPa[a] | $\rho_{2,\Phi=\Phi_0}$ / kg·m$^{-3}$ | $10^6$ ($\rho_{2,\Phi=\Phi_0}$ - $\rho_{Span}$)/$\rho_{Span}$ | $10^6$ $\varepsilon_\rho$ |
|---|---|---|---|---|---|---|---|---|
| 249.959 | 2.00764 | 27.462 | -29 | 250.009 | 1.98590 | 27.153 | -90 | 52.4 |
| 249.999 | 2.00864 | 27.471 | -27 | 250.072 | 2.00410 | 27.401 | 36 | -54.1 |
| 249.999 | 4.99450 | 69.469 | 42 | 250.018 | 4.66921 | 64.839 | 0 | 35.9 |
| 249.958 | 10.07050 | 141.612 | -6 | 250.009 | 9.99092 | 140.447 | -75 | 59.2 |
| 249.960 | 15.27000 | 211.588 | 29 | 249.976 | 16.00491 | 220.862 | 53 | -20.6 |
| 249.961 | 19.85710 | 266.561 | 10 | 250.016 | 20.30953 | 271.503 | 6 | 3.4 |
| 274.967 | 2.01164 | 24.837 | -13 | 274.983 | 1.97856 | 24.414 | -463 | 385.1 |
| 274.966 | 10.06630 | 125.016 | -20 | 274.988 | 10.04522 | 124.741 | -65 | 38.9 |
| 274.968 | 19.87500 | 235.358 | 15 | 274.994 | 20.00972 | 236.689 | 13 | 1.8 |
| 299.975 | 2.45700 | 27.684 | 34 | 299.929 | 1.96301 | 22.115 | 159 | -106.8 |
| 299.970 | 4.92558 | 55.515 | 9 | 299.980 | 3.99811 | 45.063 | -143 | 129.9 |
| 299.971 | 10.97430 | 122.253 | 27 | 299.986 | 9.99571 | 111.695 | 92 | -55.0 |
| 299.972 | 10.97440 | 122.253 | 21 | 299.962 | 10.82772 | 120.696 | 132 | -94.9 |
| 299.971 | 13.85470 | 152.560 | -29 | 299.952 | 13.98960 | 153.961 | -24 | -3.6 |
| 299.973 | 13.85490 | 152.561 | -27 | 299.968 | 14.00345 | 154.120 | 150 | -151.5 |
| 299.973 | 16.78010 | 181.982 | -12 | 299.862 | 16.01318 | 174.493 | -7 | -4.0 |
| 299.982 | 16.78760 | 182.050 | -2 | 299.990 | 15.99464 | 174.243 | 136 | -118.1 |
| 399.996 | 2.40843 | 20.140 | -92 | 400.001 | 1.99339 | 16.693 | 51 | -122.3 |
| 399.996 | 4.80372 | 39.843 | 15 | 399.981 | 3.99731 | 33.232 | -587 | 515.0 |
| 399.997 | 7.43540 | 61.038 | -32 | 400.026 | 7.94218 | 65.042 | -242 | 179.7 |
| 399.997 | 10.10050 | 81.966 | 19 | 399.982 | 9.99378 | 81.132 | -120 | 118.3 |
| 399.994 | 10.10070 | 81.970 | 37 | 400.026 | 9.93710 | 80.677 | -187 | 192.2 |
| 399.988 | 14.80710 | 117.441 | 51 | 400.028 | 13.98416 | 111.374 | 71 | -17.5 |
| 399.990 | 17.24960 | 135.043 | 37 | 400.028 | 17.71045 | 138.267 | -109 | 124.4 |
| 399.996 | 19.95270 | 153.853 | 3 | 399.980 | 19.43794 | 150.330 | 4 | -0.2 |
| 399.998 | 19.95890 | 153.899 | 33 | 400.028 | 19.93398 | 153.699 | -85 | 100.9 |

[a] Expanded uncertainties ($k = 2$): $U(T) = 4$ mK; $U(p > 3)/\text{MPa} = 75 \cdot 10^{-6} \cdot \frac{p}{\text{MPa}} + 3.5 \cdot 10^{-3}$;

$U(p < 3)/\text{MPa} = 60 \cdot 10^{-6} \cdot \frac{p}{\text{MPa}} + 1.7 \cdot 10^{-3}$.

**Table 4.** Percentage relative deviations of the experimental values of density of oxygen corrected only for the apparatus-specific effect (but not for the fluid-specific effect), $\rho_{\Phi=\Phi 0}$, and for both effects, $\rho_{\text{fluid}}$, from the density given by the reference equation of state for oxygen by Span and Wagner, $\rho_{\text{Span}}$, and the equation of state for oxygen by Schmidt and Wagner, $\rho_{\text{Schmidt}}$. Values of $\varepsilon_\rho$ obtained by using equation (17).

| $T$ / K[a] | $p$ / MPa[a] | $\rho_{\Phi=\Phi 0}$ / kg·m$^{-3}$ | $10^2 (\rho_{\Phi=\Phi 0} - \rho_{\text{Span}})/\rho_{\text{Span}}$ | $\rho_{\text{fluid}}$ / kg·m$^{-3}$ | $10^2 (\rho_{\text{fluid}} - \rho_{\text{Span}})/\rho_{\text{Span}}$ | $10^2 (\rho_{\text{fluid}} - \rho_{\text{Schmidt}})/\rho_{\text{Schmidt}}$ | $10^2 U(\rho_{\text{exp}})/\rho_{\text{exp}}$ | $10^6 \varepsilon_\rho$ |
|---|---|---|---|---|---|---|---|---|
| 250.010 | 5.691 | 91.776 | -2.949 | 94.547 | -0.019 | -0.002 | 0.083 | 83.9 |
| 250.008 | 5.002 | 79.949 | -2.973 | 82.387 | -0.014 | 0.011 | 0.092 | 84.1 |
| 250.009 | 4.001 | 63.100 | -3.010 | 65.052 | -0.009 | 0.025 | 0.111 | 84.5 |
| 250.010 | 3.001 | 46.675 | -3.044 | 48.139 | -0.003 | 0.035 | 0.143 | 84.8 |
| 250.009 | 2.000 | 30.677 | -3.077 | 31.652 | 0.004 | 0.040 | 0.207 | 85.1 |
| 250.009 | 0.999 | 15.105 | -3.129 | 15.591 | -0.009 | 0.014 | 0.398 | 86.0 |
| 273.143 | 5.868 | 84.758 | -2.738 | 87.139 | -0.005 | 0.005 | 0.080 | 84.8 |
| 273.143 | 4.999 | 71.684 | -2.764 | 73.721 | -0.002 | 0.019 | 0.092 | 85.1 |
| 273.142 | 3.999 | 56.845 | -2.795 | 58.480 | 0.001 | 0.032 | 0.111 | 85.5 |
| 273.142 | 2.999 | 42.246 | -2.822 | 43.476 | 0.007 | 0.043 | 0.144 | 85.8 |
| 273.140 | 2.000 | 27.904 | -2.852 | 28.726 | 0.011 | 0.044 | 0.208 | 86.1 |
| 273.141 | 0.999 | 13.801 | -2.890 | 14.213 | 0.004 | 0.026 | 0.403 | 86.7 |
| 293.077 | 5.524 | 73.159 | -2.598 | 75.111 | 0.001 | 0.014 | 0.085 | 85.9 |
| 293.079 | 5.000 | 66.024 | -2.612 | 67.797 | 0.002 | 0.021 | 0.092 | 86.1 |
| 293.077 | 4.001 | 52.515 | -2.640 | 53.941 | 0.003 | 0.032 | 0.112 | 86.5 |
| 293.077 | 3.000 | 39.126 | -2.666 | 40.200 | 0.006 | 0.039 | 0.145 | 86.8 |
| 293.080 | 1.998 | 25.892 | -2.693 | 26.610 | 0.007 | 0.038 | 0.211 | 87.1 |
| 293.078 | 0.998 | 12.845 | -2.720 | 13.206 | 0.009 | 0.028 | 0.408 | 87.5 |
| 324.943 | 6.055 | 71.409 | -2.383 | 73.156 | 0.005 | 0.010 | 0.079 | 87.3 |
| 324.944 | 4.999 | 58.789 | -2.410 | 60.243 | 0.003 | 0.019 | 0.094 | 87.8 |
| 324.943 | 3.999 | 46.881 | -2.432 | 48.051 | 0.004 | 0.028 | 0.114 | 88.1 |
| 324.943 | 2.986 | 34.885 | -2.458 | 35.765 | 0.001 | 0.029 | 0.148 | 88.5 |
| 324.942 | 1.999 | 23.267 | -2.482 | 23.859 | 0.000 | 0.026 | 0.214 | 89.0 |
| 324.943 | 0.999 | 11.577 | -2.525 | 11.874 | -0.021 | -0.004 | 0.416 | 90.0 |
| 349.931 | 6.122 | 66.505 | -2.256 | 68.040 | 0.000 | 0.003 | 0.080 | 88.8 |
| 349.930 | 5.001 | 54.252 | -2.282 | 55.520 | -0.004 | 0.009 | 0.095 | 89.4 |
| 349.929 | 3.999 | 43.315 | -2.302 | 44.336 | -0.004 | 0.017 | 0.115 | 89.7 |
| 349.928 | 2.985 | 32.270 | -2.330 | 33.040 | -0.012 | 0.012 | 0.150 | 90.4 |
| 349.928 | 1.998 | 21.561 | -2.346 | 22.079 | -0.008 | 0.014 | 0.218 | 90.6 |
| 349.928 | 0.999 | 10.745 | -2.403 | 11.010 | -0.046 | -0.031 | 0.424 | 92.2 |
| 374.916 | 6.145 | 61.950 | -2.145 | 63.305 | -0.004 | -0.002 | 0.081 | 90.3 |
| 374.917 | 4.999 | 50.383 | -2.167 | 51.496 | -0.006 | 0.005 | 0.096 | 90.8 |
| 374.918 | 3.996 | 40.257 | -2.173 | 41.153 | 0.005 | 0.022 | 0.117 | 90.6 |
| 374.918 | 2.980 | 30.009 | -2.188 | 30.682 | 0.007 | 0.027 | 0.153 | 90.9 |
| 374.917 | 1.987 | 19.994 | -2.198 | 20.447 | 0.014 | 0.033 | 0.223 | 90.9 |
| 374.918 | 0.999 | 10.035 | -2.214 | 10.264 | 0.016 | 0.027 | 0.431 | 91.1 |

[a] Expanded uncertainties ($k = 2$): $U(T) = 4$ mK; $U(p > 3)/\text{MPa} = 75 \cdot 10^{-6} \cdot \frac{p}{\text{MPa}} + 3.5 \cdot 10^{-3}$; $U(p < 3)/\text{MPa} = 60 \cdot 10^{-6} \cdot \frac{p}{\text{MPa}} + 1.7 \cdot 10^{-3}$.

**Table 5.** Statistical analysis of the comparison of the experimental density data for oxygen corrected for both the apparatus and the fluid effects against the two reference equations of state. AAD = absolute average deviation, Bias = average deviation, RMS = root mean square deviation, MaxD = maximum deviation.

|  | Span and Wagner [32] | Schmidt and Wagner [37] |
|---|---|---|
| Number of experimental points | 36 | 36 |
| AAD / % | 0.008 | 0.022 |
| Bias / % | 0.011 | 0.025 |
| RMS / % | 0.011 | 0.025 |
| MaxD / % | 0.046 | 0.044 |